\documentclass[10pt, conference]{IEEEtran}

\usepackage{graphicx}
\usepackage{cite}
\usepackage{amsmath,amssymb,amsfonts}
\usepackage{graphicx}
\usepackage{textcomp}
\makeatletter
\def\endthebibliography{%
	\def\@noitemerr{\@latex@warning{Empty `thebibliography' environment}}%
	\endlist
}
\makeatother
\def\BibTeX{{\rm B\kern-.05em{\sc i\kern-.025em b}\kern-.08em
		T\kern-.1667em\lower.7ex\hbox{E}\kern-.125emX}}

\usepackage{latexsym}
\usepackage{color}
\usepackage{verbatim}
\usepackage{multirow}
\usepackage{algorithm,algpseudocode}
\usepackage{lipsum}
\usepackage[flushleft]{threeparttable}
\usepackage{adjustbox}
\usepackage{booktabs}
\usepackage{makecell}
\usepackage{subfigure}



\newcommand{\bsigma}{{\boldsymbol \sigma}}

\newcommand{\bmu}{{\boldsymbol \mu}}

\algtext*{Indent}
\algtext*{EndIndent}

\IEEEoverridecommandlockouts
\begin{document}

%
\title{End-to-End Learning-Based Wireless Image Recognition Using the PyramidNet in Edge Intelligence}
%
%
%

\author{\IEEEauthorblockN{Kyubihn Lee and Nam Yul Yu}
	\IEEEauthorblockA{School of Electrical Engineering and Computer Science\\
		Gwangju Institute of Science and Technology (GIST), Gwangju, South Korea\\
		Email: l0509kb@gist.ac.kr, nyyu@gist.ac.kr}
	\thanks{This work was supported by the GIST Research Project grant funded by the GIST in 2023, and National Research Foundation of Korea~(NRF) grant funded by the Korea Government~(MSIT) (No. NRF-2022R1F1A1066143.)}}

\maketitle

\begin{abstract}
In edge intelligence, deep learning~(DL) models are deployed at an edge device and an edge server for data processing with low latency in the Internet of Things~(IoT).
In this paper, we propose a new end-to-end learning-based wireless image recognition scheme using the PyramidNet in edge intelligence. 
We split the PyramidNet carefully into two parts for an IoT device and the edge server, which is to pursue low on-device computation.
Also, we apply a squeeze-and-excitation block to the PyramidNet for the improvement of image recognition.
In addition, we embed compression encoder and decoder at the splitting point, which reduces communication overhead by compressing the intermediate feature map.
Simulation results demonstrate that the proposed scheme is superior to other DL-based schemes in image recognition, while presenting less on-device computation and fewer parameters with low communication overhead.
\end{abstract}
\begin{IEEEkeywords}
	Edge intelligence, end-to-end learning, image recognition, Internet of Things~(IoT), joint source-channel coding.
\end{IEEEkeywords}


\section{Introduction}

The development of the Internet of Things~(IoT) enables a number of IoT devices to be connected for intelligent services in offices, streets, and homes.
However, as each IoT device has limited computational power and memory space, it is challenging to provide the services with low latency~\cite{IoT_2}. 
To address the problem, \emph{edge intelligence}~\cite{IoT_3, ComMagEdge} has emerged as a new research paradigm for IoT. 
In edge intelligence, an IoT device utilizes a deep learning~(DL) model to process its data with low on-device computation and transmit with low communication overhead over wireless channels.
Then, an edge server with high computational power recovers or recognizes the data reliably using its DL model.

In edge intelligence, deep learning-based joint source-channel coding~(DL-based JSCC) schemes have been presented for reliable image recovery over wireless channels.
In \cite{DeepJSCC}, a DL-based JSCC scheme was first introduced for wireless image recovery using an end-to-end learning.
In \cite{DeepJSCC-f} and \cite{SE-DeepjSCC}, the authors improved the performance of wireless image recovery by exploiting channel feedback and attention modules for DL-based JSCC, respectively.
In \cite{DeepJSCC-bandwidth}, the authors proposed a DL-based JSCC scheme employing adaptive bandwidth transmission to allocate the bandwidth dynamically for wireless image recovery.
In \cite{DeepJSCCQ}, a DL-based JSCC scheme with a new signal constellation was proposed for wireless image recovery using a soft-to-hard quantizer. 
Recently, semantic communications~\cite{semantic_magazine} employed DL-based JSCC for text~\cite{semantic_text} and image~\cite{semantic_image} transmission.

In addition, DL-based JSCC schemes have been applied for wireless image recognition, where an edge server recognizes or classifies images through the transmitted features from IoT devices. 
Several DL-based JSCC schemes employed the ResNet~\cite{Resnet} as a feature extractor for wireless image recognition, using CIFAR-10 in \cite{access}, and CUHK03, Market-1501 and VeRi datasets in \cite{retrieval}, respectively.
In \cite{InfBottleneck}, an information bottleneck was exploited to formalize a rate-distortion trade off between compressed features and the recognition results in AWGN channel.
However, the above approaches cause large on-device computation due to huge DL model in IoT devices.

Recently, the technique of network splitting has been proposed to reduce on-device computation while maintaining reliable image recognition in edge intelligence.  
In \cite{SplitFirst}, the authors split a deep neural network to distribute computation workload across an IoT device and the edge server.
In \cite{Bottlenet++}, compression modules were embeded at the splitting point of a network to reduce communication overhead.
In \cite{SuperviseBottle}, the authors proposed a supervised compression approach to reduce communication overhead using the knowledge distillation~\cite{KnowledgeDistillation}.
In \cite{pruning}, the authors proposed a network splitting scheme together with network pruning to reduce on-device computation.
Although network splitting reduces on-device computation for an IoT device, it can degrade the performance at the edge server if noise is inserted to the splitting point. 

\begin{figure*}[ht!]
	\centering
	\includegraphics[width= 0.95\textwidth, height=0.28\textheight]{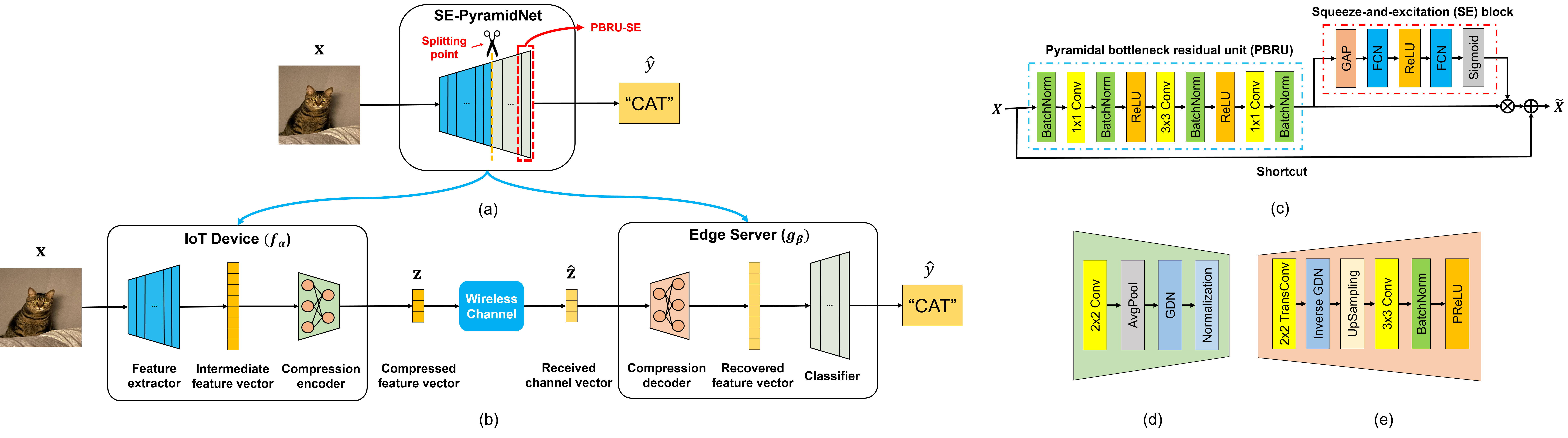}
	\hfill
	\caption{End-to-end learning-based wireless image recognition scheme using the PyramidNet~(E2E-WIR-P).
	(a) image recognition with SE-PyramidNet, (b) overview of proposed E2E-WIR-P, (c) pyramidal bottleneck residual unit with SE block~(PBRU-SE), 
	(d) compression encoder (e) compression decoder.}
	\label{Fig1}
\end{figure*}

In this paper, we propose a new end-to-end learning-based wireless image recognition scheme to improve the performance of image recognition and reduce on-device computation of an IoT device simultaneously, based on DL-based JSCC.
Unlike previous works, we adopt the PyramidNet~\cite{Pyramidnet} for reliable image recognition, which increases the feature map dimension gradually. 
Thanks to the gradual increase, the PyramidNet can overcome the performance drop that occurs when a downsampling unit is deleted from the ResNet~\cite{ResDrop}.
To reduce on-device computation, we split the PyramidNet into two parts, where the front part becomes a feature extractor on an IoT device and the rest is a classifier on an edge server, respectively. 
Also, we employ a squeeze-and-excitation~(SE)~\cite{SE} block to further improve the performance of image recognition of the PyramidNet.
In addition, we embed a compression encoder in an IoT device to reduce communication overhead and a compression decoder in an edge server to restore the original feature vector.
Contrary to the one of \cite{Bottlenet++}, we exploit asymmetric autoencoder with fewer layers on the compression encoder, to reduce on-device computation.
To the best of our knowledge, the proposed scheme is the first attempt of adopting and splitting the PyramidNet with SE blocks for wireless image recognition.
The novelty of our work is to use the split PyramidNet with SE blocks and propose new compression encoder and decoder for wireless image recognition in edge intelligence.

In simulations, we demonstrate that the proposed scheme achieves reliable image recognition with less on-device computation and fewer parameters than other DL-based schemes. 
Also, we show that the proposed scheme outperforms other DL-based schemes in image recognition with a small number of transmitted symbols, which results in low communication overhead for wireless transmission. 
In particular, it turns out that the proposed scheme is superior to the DL-based scheme of \cite{Bottlenet++} with network splitting in image recognition, while presenting less on-device computation. 
As a consequence, the proposed scheme can be suitable for reliable and efficient wireless image recognition in edge intelligence. 

\emph{Notation:}
$(\cdot)^T$ and $(\cdot)^*$ denote the transpose and the conjugate transpose of a vector, respectively.  
${\mathbf x}\sim\mathcal{CN}(\bmu,\bsigma)$ is a random vector following the circularly symmetric complex Gaussian distribution with mean $\bmu$ and covariance $\bsigma$.

\section{System model}
In this paper, we consider wireless image recognition, where an IoT device transmits the compressed feature of an image through a wireless channel, and an edge server then tries to recognize the image from the received feature. 
In the IoT device, an input image ${\mathbf x}\in\mathbb{R}^N$ is mapped to its compressed feature vector ${\mathbf z}= (z_1, \cdots, z_B)^T \in\mathbb{C}^B$ through an encoding function $f_\alpha: \mathbb{R}^N \rightarrow \mathbb{C}^B,$
where $N$ is the dimension of the input image, $\alpha$ is a trainable parameter set in the IoT device, and $B$ is the number of channel input symbols.
In wireless transmission, we impose a constraint on the average transmit power, i.e. $ \frac{1}{B}\sum_{i=1}^{B} {|z_i|}^2\leq P$.

In this system model, we assume that the compressed feature vector ${\mathbf z}$ is transmitted over the Rayleigh fading channel, where the transmitted symbols of ${\mathbf z}$ experience the same channel gain. 
The received vector in the edge server is
\begin{equation}\label{system_model}
\hat{\mathbf z}  = h{\mathbf z} + {\mathbf n},
\end{equation}
where $h\sim\mathcal{CN}(0,\sigma_h^2)$ is a channel coefficient and ${\mathbf n}\sim\mathcal{CN}({\bf 0},\sigma_n^2{\bf I})$ is the additive white Gaussian noise~(AWGN).
In (\ref{system_model}), if $h=1$, it implies the AWGN channel. 

We assume that the edge server knows the channel coefficient $h$~\cite{DeepJSCCQ, retrieval}, to investigate the performance of proposed scheme with no impact of channel estimation error.
Then, the edge server performs channel equalization by $\tilde{\mathbf z} = \frac{h^*}{|h|^2}\hat{\mathbf z}.$
In the edge server, the equalized channel vector $\tilde{\mathbf z}$ is mapped to a prediction $\hat{y}\in\mathbb{R}$ through a decoding function $g_\beta : \mathbb{C}^B \rightarrow \mathbb{R},$
where $\beta$ is a trainable parameter set in the edge server.
Finally, the edge server recognizes the image observed by the IoT device with the prediction $\hat{y}$.  
A brief overview of our system model is illustrated in Fig. \ref{Fig1}(b).

\section{Proposed Network architecture}\label{proposed network architecture}

This section describes the details of a proposed end-to-end learning-based wireless image recognition scheme using the PyramidNet~(E2E-WIR-P) in edge intelligence.
 
\subsection{SE-PyramidNet}
We adopt the PyramidNet~\cite{Pyramidnet} to improve the performance of wireless image recognition with fewer parameters and less on-device computation than the ResNet~\cite{Resnet}.
In the ResNet, it is known in \cite{ResDrop} that removing the downsampling unit of the network deteriorates the performance of image recognition.  
To overcome this issue, the PyramidNet gradually increases the feature map dimension through all residual units, in order to equally distribute the impact of downsampling unit across all residual units.
For this reason, it has been reported in \cite{Pyramidnet} that the PyramidNet shows better performance of image recognition with fewer parameters than the ResNet. 

In the PyramidNet, we use the pyramidal bottleneck residual unit~(PBRU)\footnote{The code is from \emph{https://github.com/dyhan0920/PyramidNet-PyTorch}.}, proposed in \cite{Pyramidnet}.
The PBRU consists of 3 convolution layers, 4 batch normalization~(BN) layers and 2 rectified linear unit~(ReLU) layers as in Fig. \ref{Fig1}(c). 
The feature map dimension~(FMD) of the 3$\times$3 convolution layer at the $k$th PBRU is 
\begin{equation*}
D_k = \lfloor18 +\omega(k-1)/R\rfloor,
\end{equation*}
where $R$ is the total number of PBRUs and $\omega$ is a widening factor. 
At $k=1$, the first 1$\times$1 convolution layer changes the initial FMD of 16 to $D_1=18$. 
If $k>1$, it reduces the FMD from $4D_{k-1}$ to $D_k$. 
The last 1$\times$1 convolution layer increases the FMD from $D_k$ to $4D_k$, which is the dimension of intermediate feature vector after the $k$th PBRU. 

Furthermore, we employ the squeeze-and-excitation~(SE) block~\cite{SE} to further improve the performance of image recognition. 
The SE block\footnote{The code is from \emph{https://github.com/moskomule/senet.pytorch}.}
is composed of a global average pooling~(GAP) layer, 2 fully connected neural network~(FCN) layers, ReLU and sigmoid layers as in Fig. \ref{Fig1}(c).
The first FCN of SE block at the $k$th PBRU maps $4D_k$ neurons to $D_k/4$ neurons, and the other FCN performs the inverse process of the first FCN. 
Although \cite{SEPyramid} removes the first two ReLU layers from PBRU, we maintain the layers for nonlinearity of each PBRU in embedding the SE block.
Through simulations, we observed that the PyramidNet employing SE blocks, referred to as `SE-PyramidNet~(SE-PN)', increases the top1 and top5 accuracies of the PyramidNet alone by 1$\sim$3\% with additional on-device computation of about 0.01 giga multiply-accumulate operations~(GMACs), where $R=54$ and $\omega=120$ for the PyramidNet.


\subsection{Network Splitting}
As illustrated in Fig. \ref{Fig1}(b), we split the SE-PyramidNet to distribute the computation and the parameters of SE-PyramidNet across an IoT device and an edge server. 
In network splitting, the front part is a feature extractor in the IoT device, while the rest is a classifier in the edge server. 
Since the FMD of each layer in the front is narrower than in the rest, the IoT device enjoys low on-device computation with a small number of parameters from the front part.

To balance the performance of image recognition and on-device computation in an IoT device, it is crucial to choose a splitting point of the SE-PyramidNet carefully.
If the front part of the split SE-PyramidNet increases, the image recognition accuracy improves as the intermediate feature from the IoT device gets more precise.
In this case, however, the size of DL model in the IoT device also increases, causing large on-device computation. 
In Sec. IV-B, we will investigate the recognition accuracy over on-device computation by varying the splitting point of the SE-PyramidNet.

\subsection{Compression Encoder and Decoder }
For compression encoder and decoder, we employ autoencoder architecture, which is illustrated in Fig. \ref{Fig1}(d) and \ref{Fig1}(e). 
They have asymmetric structure with fewer layers on the compression encoder to reduce on-device computation for an IoT device. 
Similar to \cite{Bottlenet++}, we use convolution layers to design compression encoder and decoder. 
Instead of batch normalization~(BN), we employ the generalized divisive normalization~(GDN)~\cite{GDN}, which is shown to be effective in compression of images~\cite{GDN_example}.

In the compression encoder, we use a single 2$\times$2 convolution layer and an average pooling~(AvgPool) layer with GDN to reduce the dimension of the intermediate feature vector. 
The number of channels in the 2$\times$2 convolution layer is adaptively changed by the size $B$ of the transmitted feature vector.
The AvgPool layer then reduces the spatial dimensionality of the intermediate feature map by a factor of 4, which results in $2B$ real values. 
The GDN layer captures the spatial information of the intermedate feature map for nonlinear transformation. 
The last normalization layer makes an input channel vector by mapping $2B$ real values into $B$ complex values, and normalizes it to satisfy the average power constraint $P$ for wireless transmission.

At an edge server, the compression decoder performs transposed convolution~(TransConv) and inverse GDN at first, which are the inverse processes of their counterparts in compression encoder. 
Then, upsampling and convolution layers proceed to recover the dimension of the original feature map. 
Finally, the decoder recovers the intermediate feature vector through the BN layer and the parametric ReLU~(PReLU). 
 

\subsection{Training Strategy}\label{training_strategy}
A direct end-to-end learning from feature extractor to classifier would suffer from the slow convergence~\cite{Bottlenet++}. 
Thus, we train the SE-PyramidNet alone and then train the compression encoder and decoder with the pretrained SE-PyramidNet. 

We consider the SE-PyramidNet in which the widening factor is $\omega=120$ and the total number of PBRU-SEs is $R=54$.
For training the SE-PyramidNet, we use the stochastic gradient descent~(SGD) optimizer with the learning rate of 0.025, and the $L_2$ regularization weighted by $5\times 10^{-4}$ for 300 epochs with the batch size of 32.
The cross-entropy~(CE) loss $l_{ce}$ is used as a loss function, i.e., $l_{ce} = -\sum_{c=1}^{C} y_c\log(p_c)$, 
where $C$ is the number of classes, $y_c$ is a target label for a training example $c$, and $p_c$ is the probability of prediction for the example $c$.
The learning rate is reduced by one-tenth at 150 and 225 epochs, respectively. 

We then train the compression encoder and decoder with the pretrained SE-PyramidNet using the CE loss. 
We use the SGD optimizer with the learning rate of 0.01 and the $L_2$ regularization weighted by $5\times 10^{-4}$ for 160 epochs with the batch size of 64.
We reduce the learning rate by one-tenth at 80 and 120 epochs, respectively. 
Finally, we train the whole E2E-WIR-P following the training strategy used for training the compression encoder and decoder. 
\begin{figure*}[th!]
	\centering
	\subfigure[]{\includegraphics[width=0.33\textwidth, height=0.22\textheight]{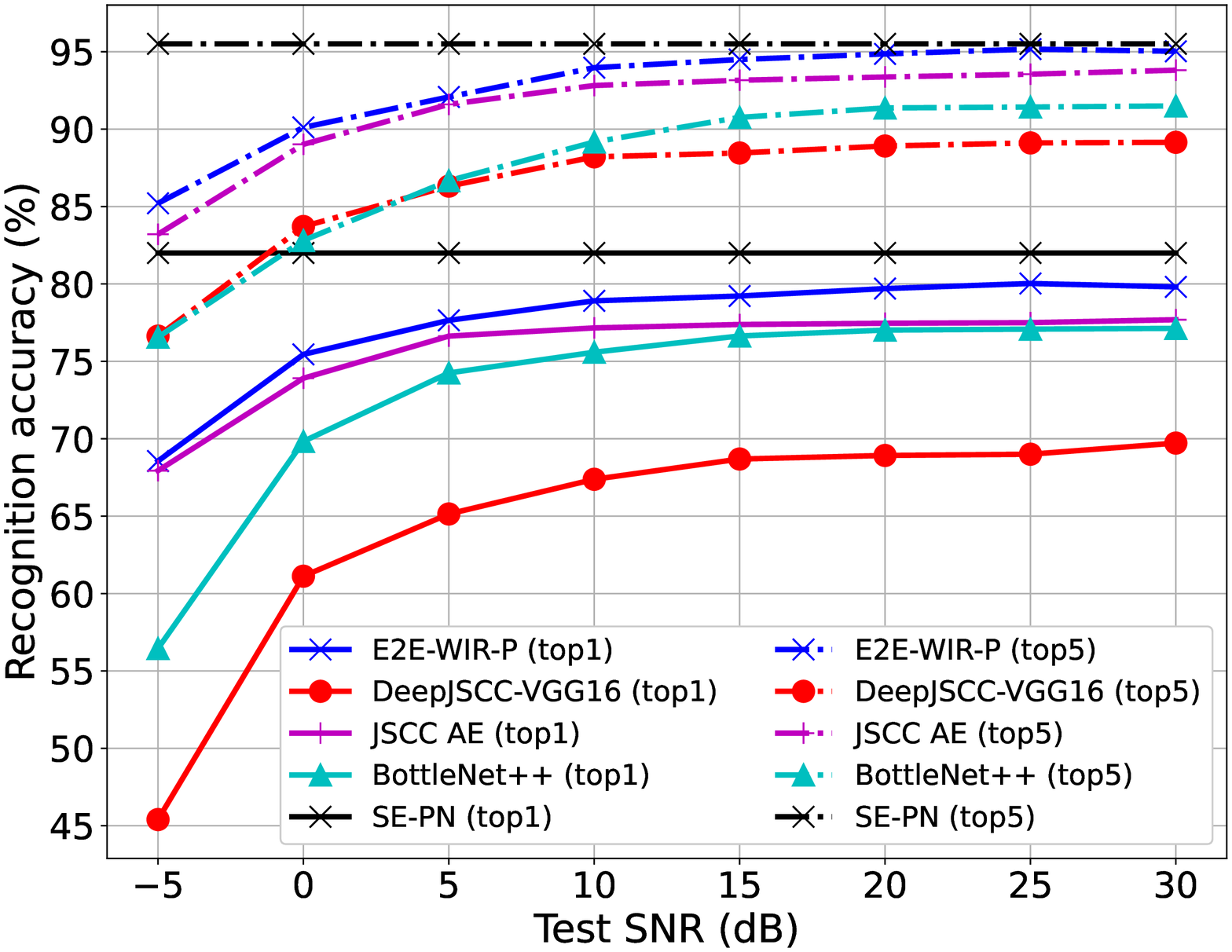}}\label{SNR}
	\hfil
	\subfigure[]{\includegraphics[width=0.33\textwidth, height=0.22\textheight]{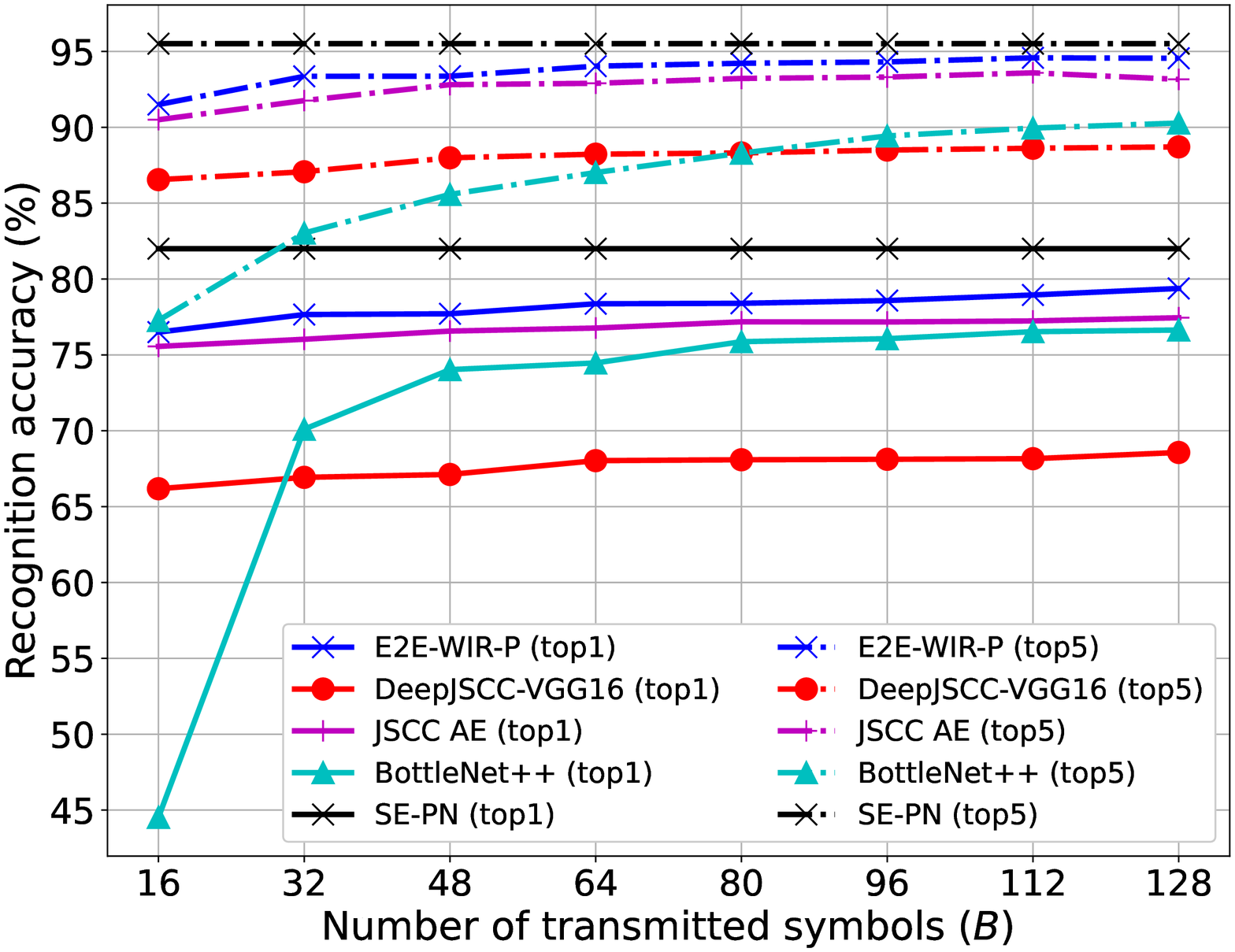}}\label{Bandwidth}
	\caption{Top1 (solid line) and top5 (dashdot line) accuracies of image recognition in Rayleigh fading over (a) test SNRs with $B=128$ and (b) the number of transmitted symbols with SNR = 15dB.}
	\label{Fig4}
\end{figure*}

\begin{figure*}[th!]
	\centering
	\centering
	\subfigure[]{\includegraphics[width=0.33\textwidth, height=0.22\textheight]{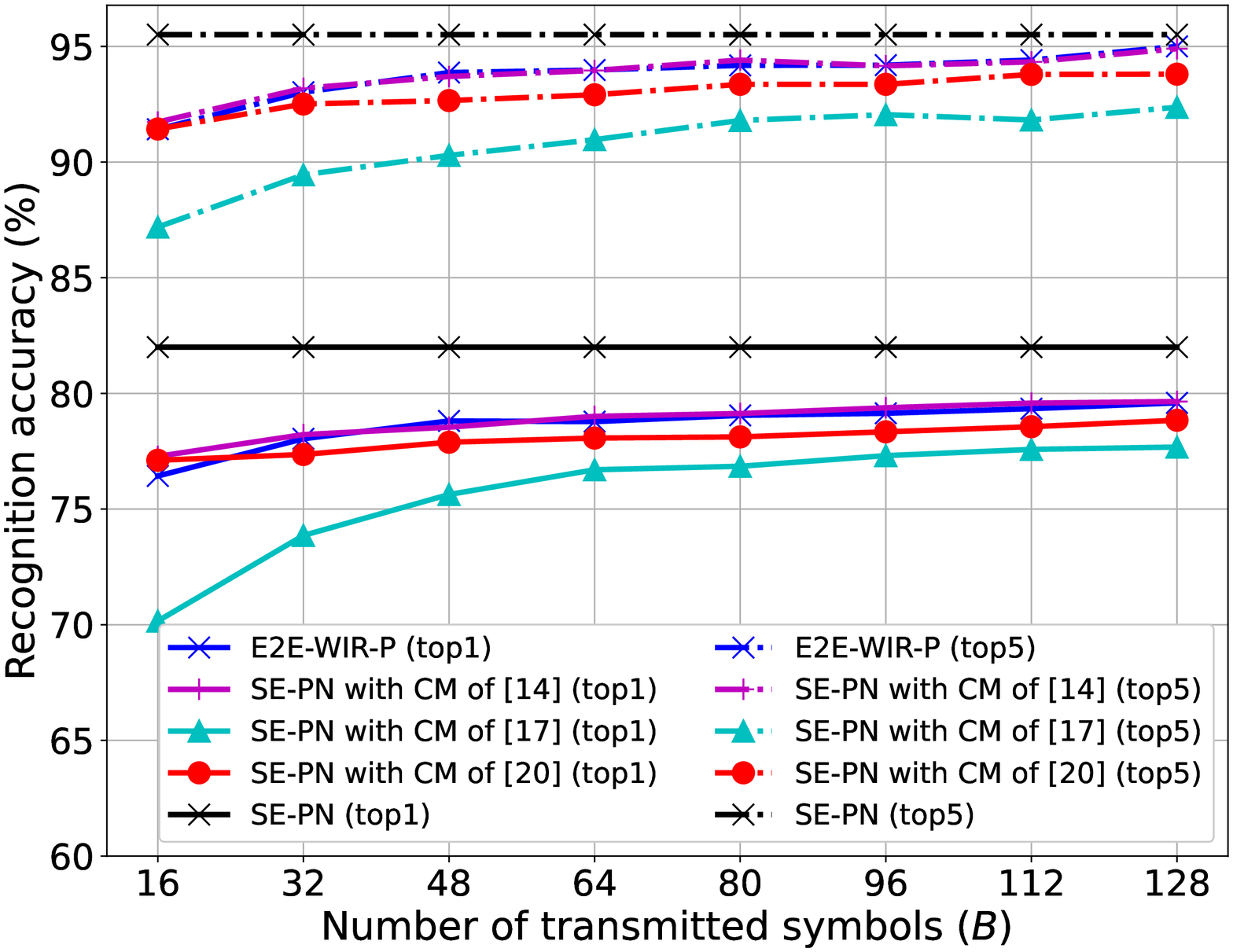}}
	\hfil
	\subfigure[]{\includegraphics[width=0.33\textwidth, height=0.22\textheight]{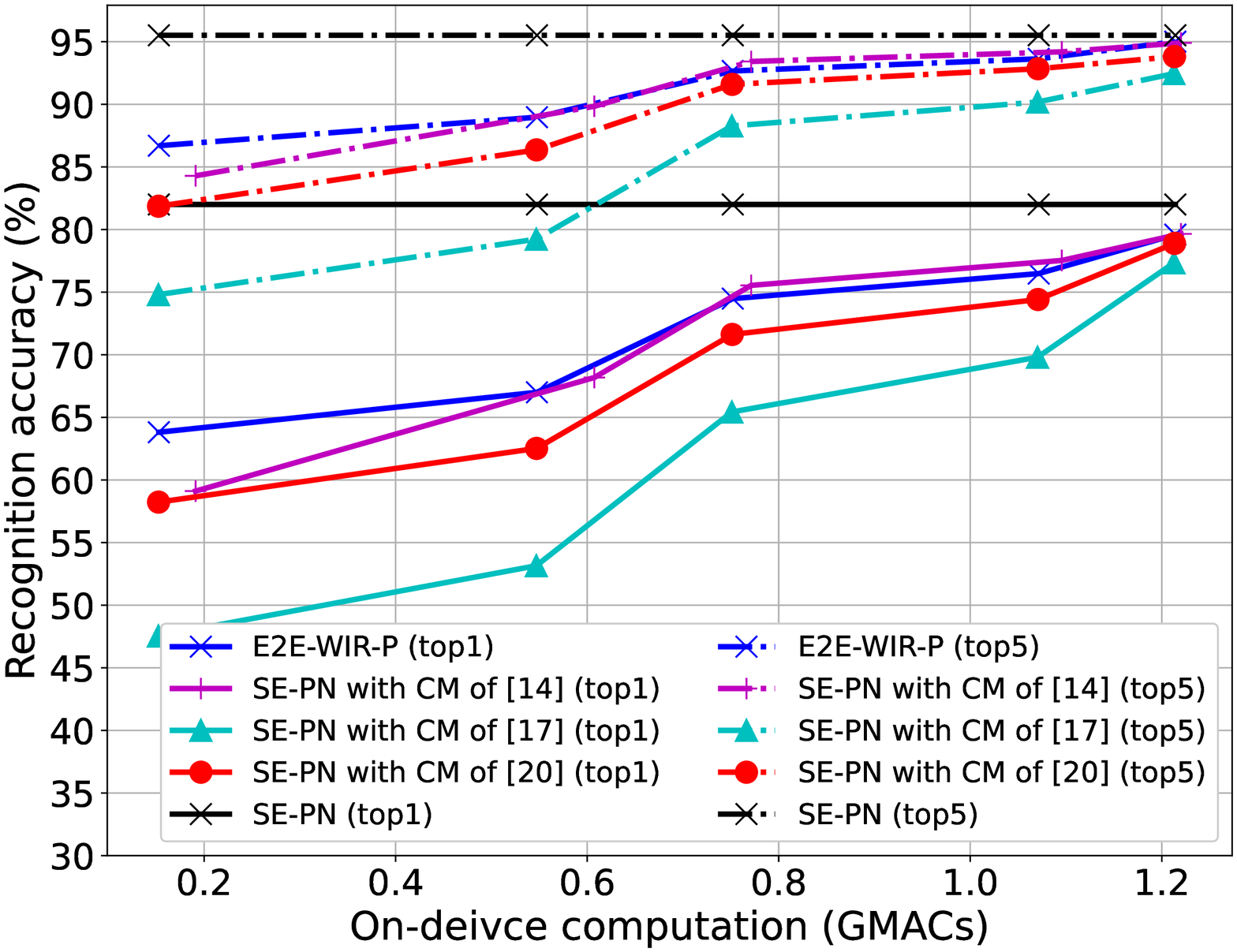}}
	\caption{Top1 (solid line) and top5 (dashdot line) accuracies of SE-PNs with different compression methods (CM) in Rayleigh fading over (a) the number of transmitted symbols and (b) on-device computation, where SNR = 15dB.}
	\label{Review1_2}
\end{figure*}
\section{Experiments}\label{Experiment}

\subsection{Experimental Setup}\label{experiment_setup}
We conduct experiments with the CIFAR100 dataset that consists of $32\times32$ RGB and labeled images coming from 100 classes. 
We use 50,000 and 10,000 images for training and test datasets, respectively. 
All experiments are implemented by using Python with the PyTorch framework on an Intel(R) Xeon(R) Gold CPU with two GeForce RTX 3090s.
We consider wireless transmission over the Rayleigh fading channel, where the signal to noise ratio~(SNR) is evaluated by SNR = $10\log_{10}\left(\frac{P\sigma_h^2}{\sigma_n^2}\right)$ in dB with $P=1$ and $\sigma_h^2=1$.
For all experiments, training and test SNRs are the same.

To evaluate the performance of image recognition, we measure top1 and top5 accuracies of the test dataset images.
The top-$k$ accuracy measures the proportion of cases that the correct label is present within the $k$ labels having the highest predicted probabilities by the model.
In addition, we compute multiply-accumulate operations~(MACs), instead of computation time, to measure on-device computation independent of GPU parallel processing mechanisms. 
We also count the number of trainable parameters in an IoT device. We use \emph{get$\_$model$\_$complexity$\_$info} function of \emph{ptflops} in the Pytorch framework to compute MACs and the number of trainable parameters.

In our simulations, the following DL-based wireless image recognition schemes are used for performance comparison.
\begin{itemize}
	\item \textbf{JSCC AE}~\cite{retrieval}: The ResNet50 without network splitting is used for feature extraction. 
	JSCC encoder and decoder compress and recover the intermediate feature vector, respectively, where they consist of FCN layers, BN layers and a Leaky ReLU layer.  
	\item \textbf{BottleNet++}~\cite{Bottlenet++}: The ResNet50 is split after the second last convolution layer, which was used in Fig. 2 of \cite{Bottlenet++}.  
	Compression modules are embeded at the splitting point, where they consist of convolution layers, BN layers and activation functions. 
	In our simulations, BottleNet++ indicates the whole end-to-end architecture including the compression modules and the split network. 
	\item \textbf{DeepJSCC-VGG16}~\cite{pruning}: The VGG16 is split after its 4th pooling layer, where the front part is for a feature extractor and the rest part is for a classifier, which was proposed in Fig. 4(b) of \cite{pruning}. 
	Compression encoder and decoder are employed at the splitting point to reduce communication overhead.
	For a fair comparison, no pruning strategy is used in our simulations. 
	\item \textbf{SE-PyramidNet~(SE-PN)}: As a benchmark, we examine the performance of SE-PN, illustrated in Fig. \ref{Fig1}(b).
	Since the SE-PN has no splitting, no compression and no wireless channel, 
	the benchmark performance gives a guideline for the highest accuracy of image recognition that can be achieved by E2E-WIR-P.
\end{itemize}

For all experiments, E2E-WIR-P is based on the SE-PN with $R=54$. 
We observed from numerical results that if the splitting point is set after the 45th PBRU, 
E2E-WIR-P achieves the recognition accuracy with at most 3\% loss from the SE-PN for sufficiently high SNR and large $B$.
So, E2E-WIR-P sets the splitting point after the 45th PBRU in the experiments of Fig. \ref{Fig4}, Fig. 3(a) and Table I.

\subsection{Performance Evaluation}\label{Evaluation}
Fig. \ref{Fig4}(a) depicts top1 and top5 accuracies of image recognition over Rayleigh fading channels as a function of test SNR, where $B=128$.
In Fig. \ref{Fig4}(a), our E2E-WIR-P is superior to the other DL-based schemes in image recognition over all SNR ranges. 
Also, only E2E-WIR-P reaches 80\% and 95\% of top1 and top5 accuracies, respectively.
In particular, if SNR$\ge20$dB, our E2E-WIR-P nearly reaches the top5 accuracy of the SE-PN and achieves the top1 accuracy with at most 2\% loss,
despite the use of network splitting with compression encoder and decoder.

Fig. \ref{Fig4}(b) shows top1 and top5 accuracies of image recognition over Rayleigh fading channels for a range of the number of transmitted symbols $B$, where SNR = 15dB. 
In Fig. \ref{Fig4}(b), E2E-WIR-P outperforms the other schemes in image recognition over a wide range of $B$.
Fig. 2(b) reveals that as long as $B\geq 64$, E2E-WIR-P are almost stable over $B$ with 3.5\% and 0.5\% losses of top1 and top5 accuracies, respectively, from those of the SE-PN. This suggests that E2E-WIR-P has little influence from the number of transmitted symbols, thanks to the proposed compression encoder and decoder, which allows low communication overhead.
We performed the experiments of Fig. \ref{Fig4} for AWGN channel, which showed that the performance trends are similar to Fig. \ref{Fig4}.

In Fig. 3, we compare the recognition accuracies of E2E-WIR-P with those of SE-PN using other compression methods~(CM) in [14], [17], [20].
In Fig. 3, E2E-WIR-P outperforms the SE-PN with CM of [17] and [20] over a wide range of $B$ and on-device computation.
Only the accuracies of SE-PN with CM of [14] are similar to those of E2E-WIR-P, but the SE-PN with CM of [14] needs more parameters than E2E-WIR-P to train the compression encoder based on FCN.
Fig. 3 reveals that the CM of E2E-WIR-P can reduce the number of transmitted symbols effectively with the backbone network of SE-PN by preserving the recognition accuracies with fewer parameters, which shows the superiority of our compression encoder and decoder.

\begin{figure}[t]
	\centering
	\includegraphics[width=0.38\textwidth, height=0.22\textheight]{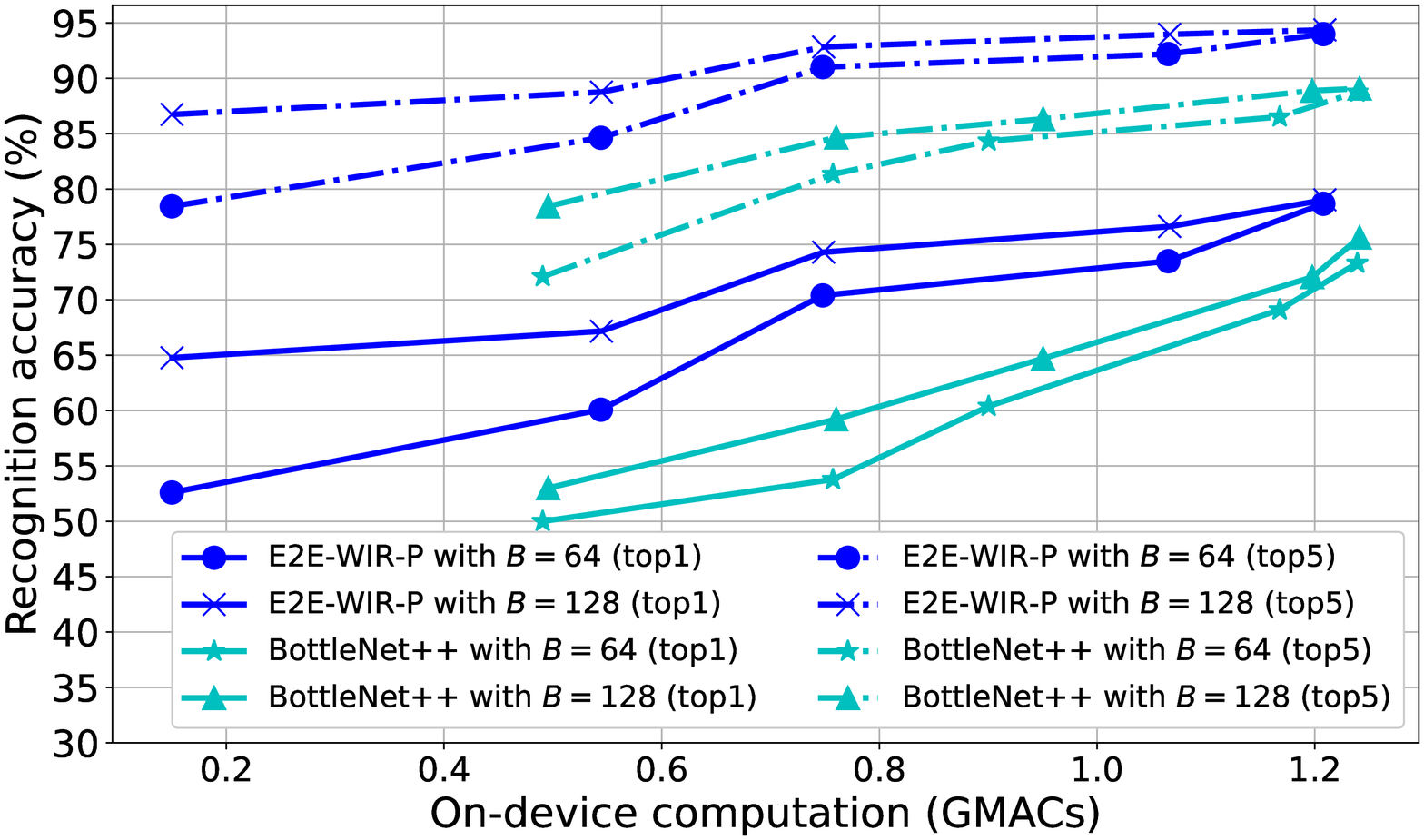}
	\caption{Top1~(solid line) and top5~(dashdot line) accuracies of E2E-WIR-P and BottleNet++ over on-device computation in Rayleigh fading, where SNR = 15dB, $B=64$ and $128$.}
	\label{Split}
\end{figure}

Since E2E-WIR-P and BottleNet++ have network splitting and show good performance in Fig. \ref{Fig4},
it is worth comparing their recognition accuracies over on-device computation by varying their splitting points. 
Fig. \ref{Split} depicts top1 and top5 accuracies of E2E-WIR-P and BottleNet++ over on-device computation, where SNR = 15dB, $B=64$ and $128$.
In Fig. \ref{Split}, if the splitting point is set backward, E2E-WIR-P and BottleNet++ with $B=64$ reach their recognition accuracies for $B=128$, respectively.
In particular, it turns out that with the same on-device computation, the recognition accuracies of E2E-WIR-P with $B=64$ are higher than those of BottleNet++ with $B=128$,
which demonstrates that E2E-WIR-P achieves reliable image recognition with less communication overhead.
Also, our E2E-WIR-P reaches the top1 accuracy of 75\% with only about a half of the on-device computation of BottleNet++.
In the end, E2E-WIR-P enables more reliable image recognition with less on-device computation and less communication overhead than BottleNet++.

\subsection{Complexity Analysis}
\begin{table*}[!th]
	\footnotesize
	\label{On-device}
	\caption{On-device computation and the number of trainable parameters in an IoT device, where $B=128$.}
	\centering
	\begin{threeparttable}
		\begin{tabular}{c|c|c|c|c|c|c}
			\toprule
			\makecell[c]{Networks} & \makecell[c]{E2E-\\WIR-P}          & \makecell[c]{JSCC\\AE}   & {BottleNet++}      & \makecell[c]{DeepJSCC-\\VGG16}      \\
			\midrule
			\makecell[c]{On-device \\computation (GMACs)}   & 1.211  & 1.311 & 1.243 & 0.288\\	
			\midrule
			\makecell[c]{Number of \\ trainable parameters (M)}       &  5.374  & 29.009 & 21.697 & 10.001    \\
			\bottomrule
		\end{tabular}
	\end{threeparttable}
\end{table*}
Table I compares the number of trainable parameters and on-device computation of an IoT device for the DL-based wireless image recognition schemes, where the number of transmitted symbols is $B=128$. 
In Table I, our E2E-WIR-P has less on-device computation and fewer parameters than JSCC AE and BottleNet++. 
Although the on-device computation is larger than that of DeepJSCC-VGG16, our E2E-WIR-P shows much higher accuracies in image recognition, as shown in Fig. \ref{Fig4}.

The experimental results of this section demonstrate that our E2E-WIR-P can be a suitable option for wireless image recognition in edge intelligence. 
It is remarkable that the SE-PyramidNet-based end-to-end learning can achieve more reliable and efficient image recognition than the VGG and the ResNet-based ones in edge intelligence.

\section{Conclusion}
In this paper, we have proposed a new end-to-end learning-based image recognition scheme for edge intelligence. 
We adopted and split the SE-PyramidNet for the first time in edge intelligence.  
Also, we embeded compression encoder and decoder at the splitting point, to reduce communication overhead. 
Simulation results demonstrated that the proposed scheme is superior to other deep learning-based schemes in image recognition, while presenting less on-device computation and fewer parameters in an IoT device with low communication overhead.
Therefore, the proposed scheme can be applied for reliable and efficient wireless image recognition in edge intelligence.
A further study will be fruitful for investigating the robustness of the proposed scheme in different wireless channel environments, e.g., multipath fading.

\bibliographystyle{IEEEtran}
\bibliography{mybibfile}

\end{document}